\documentclass{article}
\pdfoutput=1 
\usepackage[normalem]{ulem}
\usepackage{bm}
\usepackage[tbtags]{amsmath}
\usepackage{amsfonts}
\usepackage{mathtools}
\usepackage{amssymb}
\usepackage[T1]{fontenc}
\usepackage[utf8]{inputenc}
\usepackage{lmodern}
\usepackage[english]{babel}
\usepackage{latexsym}
\usepackage{braket}
\usepackage{physics}
\usepackage{slashed}
\usepackage{verbatim}
\usepackage{graphicx}
\usepackage{units}
\usepackage{booktabs}
\usepackage{appendix}
\usepackage{cite}
\usepackage{float}
\usepackage[textheight=\textheight,textwidth=\textwidth,bindingoffset=1cm,vcentering]{geometry}

\usepackage{jcappub}

\usepackage{color}
\usepackage{amsfonts}
\usepackage{graphicx}
\usepackage{amssymb}

\usepackage{mathrsfs}

\usepackage{mathtools}
\usepackage{ulem}
\usepackage{bm}
\usepackage{xcolor}
\RequirePackage{doi}
\usepackage{hyperref}

\renewcommand{\thesection}{\arabic{section}}

\def\laq{~\raise 0.4ex\hbox{$<$}\kern -0.8em\lower 0.62ex\hbox{$\sim$}~}
\def\gaq{~\raise 0.4ex\hbox{$>$}\kern -0.7em\lower 0.62ex\hbox{$\sim$}~}

\RequirePackage{doi}
\usepackage{titlesec}								
\titleformat{\chapter}
{\filright\normalfont\huge\bfseries}
{\thechapter}
{5pt}
{}

\date{}

\def\beq{\begin{equation}}
	\def\eeq{\end{equation}}
\def\bea{\begin{eqnarray}}
	\def\eea{\end{eqnarray}}

\def \pa {\partial}
\def \ra {\rightarrow}
\def \le {\left}
\def \ri {\right}

\def \b {\beta}

\def \e {\epsilon}

\def \L {\Lambda}

\def \r {\rho}

\def \om {\omega}
\def \Om {\Omega}
\def \f {\phi}

\def \p {\psi}

\def \t {\tau}

\def \mc {\mathcal}

\def \pa {\partial}
\def \ra {\rightarrow}

\def \ap {\dot{a}}
\def \app {\ddot{a}}

\def \lag {\langle} 
\def \rag {\rangle}

\def \ve {\mathbf}
\def \br {\langle}
\def \ke {\rangle}
\title{On Adiabatic Renormalization with a Physically Motivated Infrared Cut-Off}

\author[a]{Chiara Animali,}
\author[a]{Pietro Conzinu,}
\author[a]{and Giovanni Marozzi}

\affiliation[a]{Dipartimento di Fisica, Universit\`a di Pisa, Largo B. Pontecorvo 3, 56127 Pisa, 
Italy,\\
and Istituto Nazionale di Fisica Nucleare, Sezione di Pisa, Italy}

\emailAdd{chiara.animali@phd.unipi.it}
\emailAdd{pietro.conzinu@phd.unipi.it}
\emailAdd{giovanni.marozzi@unipi.it}

\abstract{
We introduce a new approach to renormalize physical quantities in curved space-time by  adiabatic subtraction. We use a comoving infrared cut-off in defining the adiabatic counterpart of the physical quantity under consideration, building on the fact that the adiabatic approximation is ill-defined in the infrared tail of the spectrum. 
We show how this infrared cut-off should be used to obtain a completely well-defined renormalization scheme and how it is  fundamental to avoid unphysical divergences that can be generated by a pathological behavior of the adiabatic subtraction extended to the infrared tail.
The infrared cut-off appears as a new degree of freedom introduced in the theory and its actual value has to be consistently fixed by a physical prescription. As an example, we show how such degree of freedom can be set to obtain the correct value of the conformal anomaly in the symptomatic case of an inflationary model with gauge fields coupled to a pseudo-scalar inflaton.
}

\makeatletter
\gdef\@fpheader{}
\makeatother
\begin{document}
\maketitle

\section{Introduction}\label{Introduction}
Inflation, an early phase of accelerated expansion, was originally proposed to solve the problems associated with the Hot Big-Bang standard theory
\cite{Guth:1980zm,Linde:1981mu}. It was then realized that inflation could also generate a nearly scale invariant spectrum of scalar and tensor
fluctuations \cite{Mukhanov:1981xt,Starobinsky:1979ty}, already in its minimal version with only  one real scalar field.

The simplicity of the inflationary idea and its success in providing a theory of initial conditions made inflation, from one of the possible scenarios of the early Universe, one of the most recognized and accepted candidates, leading to a continuous investigation of different inflationary models and to the study of their predictions and phenomenology (see, for instance, \cite{Martin:2013tda}, for an exhaustive collection of inflationary models). Due to its relevance, inflation is also a natural playground to study aspects of quantum fields in curved space-times.

Within the framework of the inflationary paradigma, it is well-known that correlation functions (or in general bi-linear observables) of quantum fields on a curved background suffer from divergences. In general,
the presence of ultraviolet (UV) divergences due to fluctuations on arbitrary short scales is a common aspect of quantum field theory \cite{Collins:1984xc}. 
In flat space and for free theories, infinities can be removed by normal ordering, namely, by subtracting the expectation value of the vacuum energy. This is legitimate, indeed this contribution is not observable.
Differently, in curved space-time such divergences cannot be easily cured as in
flat space \cite{Birrell:1982ix},
since the vacuum is not unambiguously defined. 

Among the several renormalization schemes proposed,
the one most commonly used in the context of inflationary cosmology is the \textit{adiabatic renormalization} \cite{Zeldovich:1971mw, Parker:1974qw,Fulling:1974zr, Bunch:1980vc, Anderson:1987yt}.
The adiabatic procedure to renormalize divergent quantities in curved space-times is based on subtracting the expectation value of such quantities associated with the adiabatic vacuum, which is the vacuum that minimizes the creation of particles due to the presence of a time-dependent metric. The advantage of the adiabatic renormalization relies on the fact that on one hand its physical interpretation is clear and on the other hand its implementation is straightforward. 

From its introduction \cite{Zeldovich:1971mw, Parker:1974qw, Fulling:1974zr, Bunch:1980vc} this method has been applied in various examples with success (see \cite{Birrell:1982ix,Parker:2009uva, Fulling:1989nb} for a comprehensive review), however in recent years its correct use and its consequences on physical observables (like the CMB power spectrum) were subjects of several investigations and controversy. 
For example,
in \cite{Parker:2007ni, Agullo:2008ka}, it was argued that to properly evaluate the power spectrum of inflationary fluctuations, adiabatic subtraction should be taken into account, since such spectrum diverges at coincident points. In particular, on one hand in 
\cite{Agullo:2009vq} this idea was applied by subtracting the adiabatic term at the Hubble exit,
obtaining results that differ significantly from the standard ones \cite{Planck:2018vyg}.
 On the other hand in \cite{Durrer:2009ii} it was suggested that the right time to perform the subtraction is the end of inflation rather than the Hubble exit, in which case the impact of the adiabatic subtraction on the scalar and tensor power spectra 
is subleading.

As discussed in \cite{Durrer:2009ii}, the main issue in the adiabatic subtraction procedure is that it is ill defined when the scales of interest are stretched beyond the Hubble
horizon\footnote{See also \cite{Finelli:2007fr, Marozzi:2011da} for further criticism on the
main idea of \cite{Parker:2007ni, Agullo:2008ka, Agullo:2009vq}.}. In support of this claim, in recent literature there are cases in which the adiabatic subtraction introduces unphysical infrared (IR) divergences (see, for example, \cite{Ballardini:2019rqh, Kamada:2020jaf}) when performed over all the k-spectrum. For example, 
in \cite{Ballardini:2019rqh}  
it was shown, in the case of massless gauge fields
coupled to a pseudo-scalar
inflaton, how the adiabatic renormalization correctly removes the UV divergences but leads to IR divergences.
In particular, despite the bare values of the energy density and of the helicity of gauge fields are not divergent in the infrared, their adiabatic counterparts introduce logarithmic divergences associated to the behaviour of the adiabatic approximation for massless fields in the infrared tail.

The above points suggest how the adiabatic subtraction extended over the full IR domain leads to practical and conceptual issues.
In the following we will show how to modify the usual adiabatic renormalization procedure along this direction, introducing a comoving IR cut-off, and we will show how to fix it univocally, hence the renormalization scheme, by a physically motivated prescription  for the mentioned case of massless gauge fields coupled to a pseudo-scalar inflaton.

The paper is organized as follows. In Sec. \ref{section2} we describe the adiabatic subtraction procedure and we emphasize the proper range of application of such adiabatic subtraction. In Sec. \ref{sezione3} we describe how to extend the adiabatic renormalization scheme by inserting a comoving IR cut-off and we show how to fix such cut-off univocally by a proper physical prescription for the phenomenologically interesting case of a gauge field coupled with a pseudo-scalar inflaton by a Chern-Simons-like term. In such model, in fact, the usual adiabatic regularization method exhibits problematic aspects. 
We show how the renormalization scheme can be fixed requiring to match the proper value of the conformal anomaly of gauge fields and how this procedure leads to well-defined finite results for their energy density and helicity integrals. Finally, in Sec. \ref{sezione Conclusioni} we present our final remarks and
conclusions. 
In Appendix \ref{A}  we comment on the covariant conservation of the energy momentum-tensor when adiabatic subtraction with a comoving IR cut-off is performed, whereas Appendix \ref{B} extends the adiabatic results of Sec. \ref{sezione3} to the case of massive gauge fields.

\section{Adiabatic subtraction and the need for an infrared cut-off}
\label{section2}
Let us introduce the adiabatic regularization method by studying the pedagogical (but physically motivated) case of a 
test scalar field in a curved space-time \cite{Birrell:1982ix, Parker:2009uva}. 
We consider a Friedmann-Lemaître-Robertson-Walker (FLRW) metric
\beq
{\rm d}s^2={\rm d}t^2-a^2(t)\, {\rm d}\ve{x}^2\,,
\label{FLRWmetric}
\eeq
and a Lagrangian density given by
\beq
\mc{L}= \frac{1}{2} |g|^{1/2}\le(g^{\mu \nu} \pa_\mu \phi \pa_\nu \f -m^2 \f^2- \xi R \f^2 \ri) \,,
\eeq
with $a(t)$ the scale factor, $\xi$ a dimensionless coupling constant and $R$ the space-time scalar curvature. 
It follows that the equation of motion of the scalar field is
\beq
\le( \Box +m^2 + \xi R \ri)\f=0\,.
\label{eq 2.3}
\eeq
 In close analogy to the Minkowski case, we can proceed with the standard quantization of the scalar field, expanding the field operator as
\beq
\f(x)= \sum_{\ve{k}} \{ A_{\ve{k}} f_{\ve{k}}(x)+ A^\dagger_{\ve{k}} f^*_{\ve{k}}(x) \}\,,
\eeq
where $A^\dagger_{\ve{k}}$ and  $A_{\ve{k}} $ are the creation and annihilation operators and the mode function $f$ is given by\footnote{V represents the spatial volume of a box, the continuum limit is approached for $V \rightarrow \infty$.}
\beq
f_{\ve{k}}= (2V)^{-1/2} a(t)^{-3/2} h_k(t)e^{i \ve{k}\cdot \ve{x}}\,.
\eeq
From Eq. \eqref{eq 2.3} it follows that the rescaled mode function $h_k(t)$ satisfies the equation
\beq
\ddot{h}_k+\Om^2_k \,h_k=0\,,
\label{eq 2.6}
\eeq
where a dot denotes a derivative w.r.t. cosmic time $t$ and the frequency is given by \footnote{We explicitly show the expression of the frequency but the argument of this section is completely general.} 
\beq
\Om_k^2=\om_k^2+\sigma\,,
\eeq
with
\beq
\om_k(t)=(k^2/a(t)^2+m^2)^{1/2}\,\,,\qquad \sigma(t)=(6\xi -3/4) (\dot{a}/a)^2+(6\xi-3/2) \ddot{a}/a\,.
\eeq

The adiabatic renormalization method relies on the Wentzel-Kramer-Brillouin (WKB) approximation of the mode function $h_k$
\beq 
h_k(t)= \frac{1}{\sqrt{2 W_k(t)}} e^{-i \int W_k(t^\prime){\rm d}t^\prime}\,,
\label{eq 2.9}
\eeq
where $W_k(t)$ can be determined by inserting the WKB ansatz \eqref{eq 2.9} into the equation of motion \eqref{eq 2.6}. This leads to the non-linear differential equation
\beq
W_k(t)^2=\Om_k(t)^2-\left( \frac{\ddot{W}_k(t)}{2W_k(t)}-\frac{3 \dot{W}_k(t)^2}{4  W_k(t)^2}\right)\,,
\eeq
which in general is not possible to solve. 
However, solutions of this equation can be obtained iteratively in the approximation of a background that slowly changes in time (namely, under the condition of adiabatic expansion).
This can be pictured by introducing an adiabatic parameter $\epsilon$ that describes the slowness of the time variation of the metric, and which is thus assumed to be  $\epsilon \ll 1$, and by assigning a power of $\epsilon$ to each time derivative. In this way, the solution for $W_k(t)$ is obtained as a power series in time derivatives
\beq 
W_k(t)=W_k^{(0)}(t)+ \e\, W_k^{(1)}(t) +\cdots + \e^n\, W_k^{(n)}(t) \, ,
\eeq
where $W_k^{(n)}$ is given by iterating the recursive equation up to order $n$.
Furthermore, let us add that time derivatives can be thought as curvature derivatives, and thus the expansion in time also becomes an expansion in curvature  \cite{Birrell:1982ix,Parker:2009uva}.\footnote{This point can be better understood considering the adiabatic approximation as the right approximation to minimize particle creation in the limit where the single particle energy is larger with respect to the energy scale determined by the curvature of the space-time \cite{Birrell:1982ix,Parker:2009uva}.}
Thereby, following this procedure, the adiabatic solution for the mode function can be obtained at each adiabatic order.

At this point, the proper renormalization based on the adiabatic method is then realized performing the subtraction between a bare UV-divergent quantity and its adiabatic counterpart, up to the right adiabatic order needed to remove the UV divergences.

As a representative example, let us consider the expectation value of the energy-momentum tensor $T_{\mu \nu}$. The finite result is given by the subtraction
\beq 
\langle{T_{\mu\nu}} \rangle_\text{ren}=\langle T_{\mu\nu} \rangle _\text{bare} - \br T_{\mu\nu} \ke_\text{ad} \, ,
\eeq
where the first term on the r.h.s is the bare quantity while the second one is its adiabatic counterpart.
For this particular case of the energy-momentum tensor one should consider the adiabatic expansion up to fourth order to be able to cancel the UV divergences of the bare expectation value \cite{Birrell:1982ix, Parker:2009uva}. 

From this brief introduction, it is immediate to grasp the power of the adiabatic renormalization procedure, thanks to its intuitive physical meaning as well as to its straightforward implementation. However, let us remark that adiabatic renormalization (or, equivalently, regularization) concerns the renormalization of UV divergences, essentially because the WKB ansatz for the mode functions matches exactly the solution in the deep UV, where the space-time is well approximated by the Minkowski one.

To better understand this point, we can consider again the WKB expression in Eq. \eqref{eq 2.9}. As we can see it describes an oscillating solution, which is indeed suitable only for modes that experience a negligible gravitational interaction. In a cosmological fashion we should say that this is a good approximation only for those modes that are sub-horizon.
This last aspect is properly the main point to keep in mind. However, in the common practice, the adiabatic subtraction is generally extended also to the IR domain, despite this is not properly fair, since, as said, the adiabatic approximation is not well-defined in the IR (superhorizon) regime \cite{Durrer:2009ii}.

The argumentation is twofold, indeed let us consider
the adiabatic expansion up to a generic order $n>4$ for the energy-momentum tensor, integrating in the momentum space between $k=0$ and a UV cut-off $\Lambda$. Since each adiabatic order brings to a derivative term in the expansion, from a dimensional analysis, it follows that the adiabatic expectation value $\langle T_{\mu \nu}\rangle_\text{ad}$ will have the following general structure\footnote{We explicitly checked the general structure of Eq. \eqref{General-structure-EMT} for the model discussed in Sec. \ref{sezione3}.}
\begin{equation}
\br T\ke _\text{ad}^{(n>4)}= H^{4}\sum_{n>4} \left( c_n \left(\frac{H}{m} \right)^{n-4}+c'_n \left(\frac{H}{\Lambda}\right)^{n-4}\right)\,,
\label{General-structure-EMT}
\end{equation}
where 
$H=\dot{a}/a$ is the Hubble parameter and the coefficients $(c_n, c'_n)$ are fixed by the particular model.

From the above general expression we can note two important points. In the deep UV domain, when $\L \rightarrow \infty$, the higher order terms go to zero and we can truncate the series at the fourth adiabatic order, which is indeed the order needed to remove the UV divergences. On the other hand, the IR regime produces higher order terms involving $m$ which are increasingly relevant for $m<H$ (as we generally have), and, in particular, for $m \rightarrow 0$. Therefore, to extend the integral to such regime, to be consistent, we should include all orders in the calculation.

This seems to indicate that in order to extend the adiabatic subtraction also to the IR regime, the series cannot in general be truncated at the given order needed to remove the infinities in the UV, but it should be considered up to all orders.

Therefore, in light of the above arguments, we suggest that the procedure of adiabatic regularization should be always performed on a proper domain which excludes the IR tail of the spectrum.
Namely, the adiabatic subtraction should be considered only up to a comoving IR cut-off $c=\b a(t) H(t)$.
This IR cut-off is associated to the scale at which the adiabatic solution is not anymore a good approximation for the mode functions,  condition that happens when the modes start to feel the curvature of space-time. In other words, it is related to the horizon "exit" of modes, and the coefficient $\b$, which is a new free parameter introduced by the renormalization method, should be determined by a proper physical prescription, fully in line with the spirit of each renormalization scheme  \cite{Collins:1984xc}.
It should be noted that this IR cut-off is related only to the adiabatic counterpart, while the bare quantity is free to run in the total momentum space.


\section{Infrared cut-off and conformal anomaly matching}
\label{sezione3}

The extension of the adiabatic renormalization method here proposed has an interesting and remarkable application in the study of the energy density and helicity of gauge fields coupled with an axion-like inflaton ( see e.g. \cite{Ballardini:2019rqh, Anber:2009ua,Turner:1987bw,Garretson:1992vt,Adshead:2016iae,Sobol:2019xls,Adshead:2015pva,McDonough:2016xvu,Domcke:2018eki,Barnaby:2011vw,Domcke:2020zez, Caravano:2021bfn, Hashiba:2021gmn, Ishiwata:2021yne, Lozanov:2018kpk} for the rich phenomenology associated with axion-like inflationary models and the production of gauge fields in this context). 
As shown in \cite{Ballardini:2019rqh}, the standard adiabatic renormalization of these two quantities, despite correctly removes the divergences in the UV, also introduces unphysical IR divergences, leading to not well-defined final results.

Following for example \cite{Ballardini:2019rqh}, the Lagrangian of the model is given by
\beq\label{Lag}
\mathcal{L}=-\frac{1}{2} (\nabla \phi )^2 - V(\phi) - \frac{1}{4} (F^{\mu\nu})^2 - \frac{g\phi}{4} F^{\mu\nu} \tilde{F}_{\mu\nu}\,,
\eeq
where  $\tilde{F}^{\mu\nu}=\epsilon^{\mu\nu\alpha\beta} F_{\alpha \beta}/2= \epsilon^{\mu\nu\alpha\beta}(\partial_\alpha A_\beta-\partial_\beta A_\alpha)/2$, $\nabla$ is the covariant derivative, and the coupling constant $g$ can be expressed in terms of the axion decay constant $f$ by the relation $g=\alpha/f$, with $\alpha$ a dimensionless parameter. Finally, the background is assumed to be described by a FLRW metric as given in Eq. (\ref{FLRWmetric}).

Due to the coupling with the inflaton field $ \f$ in Eq. \eqref{Lag}, quantum fluctuations of the gauge field $A_\mu$ are amplified. 
 In this context, two interesting quantities that can be considered are the following. The first is the vacuum expectation value of the energy density of the produced gauge fields
\beq
\frac{\br\ve{E}^2+\ve{B}^2\ke}{2}=\int \frac{{\rm d} k}{(2\pi)^2 a(\tau)^4} k^2 \left[|A'_+|^2+|A'_-|^2+k^2 \le(|A_+|^2+|A_-|^2\ri)\right]\,.
\label{Energy formal}
\eeq
This
is given by the $(0,0)$ component of the associated energy-momentum tensor 
\beq
T_{\mu\nu}^{(F)}=F_{\rho\mu} F^\rho_\nu + g_{\mu\nu} \frac{\ve{E}^2 - \ve{B}^2}{2}\,,
\eeq
and it enters in the Friedmann equations in the following way 
\beq
H^2 ={1\over 3 M_p^2}\le [{\dot{\f}^2\over 2} + V(\f) +{\lag \ve{E}^2+\ve{B}^2\rag \over 2}\ri] \, , \qquad \quad \dot{H} =-{1\over 2 M_p^2}\le [\dot{\f}^2+{2\over3}\lag \ve{E}^2+\ve{B}^2\rag \ri]\,.
\eeq
The second is the so-called helicity integral (following the notation of \cite{Ballardini:2019rqh}), given by
\beq
\le\langle{\ve{E}\cdot\ve{B}}\ri\rangle=- \int \frac{{\rm d}k }{(2 \pi)^2 a(\tau)^4}k^3\frac{\pa}{\pa \t} \le( |A_+|^2-|A_-|^2 \ri)\,,
\label{Helicity formal}
\eeq
which affects the equation of motion of the pseudo-scalar field as
\beq
\ddot{\f}+3 H \dot{\f} +V_\f=g \le\langle{\ve{E}\cdot\ve{B}}\ri\rangle\,.
\eeq
In all the above expressions $\ve{E}$ and $\ve{B}$ are the electric and magnetic fields associated to the gauge field $A_\mu$ and the expectation values in Eqs. \eqref{Energy formal} and \eqref{Helicity formal} are explicitly expressed in terms of the mode functions $A_\pm$ (where the basis of circular polarization has been chosen). Moreover, a prime stands for derivative w.r.t. the conformal time $\t$ (${\rm d} \tau= {\rm d}t/a$).

 The Fourier mode functions $A_\pm$ of the gauge fields satisfy the equation of motion
\beq\label{eom_A}
\frac{{\rm d}^2 }{{\rm d} \tau^2} A_\pm(\tau,k) + \left(k^2 \mp k g \f^\prime \right) A_\pm(\tau,k)=0\,.
\eeq
Under the assumption of de Sitter expansion for the background (i.e. $a(\t)=-1/(H \t)$ with $\t<0$, $H= \text{const.}$ and $\dot{\f}= \text{const.}$), we can rewrite the above equation in terms of the constant parameter $\xi \equiv g  \f^\prime /(2 a(\t)  H)= g  \dot{\f} / ( 2 H) $. In this approximation the analytical solution of Eq.(\ref{eom_A}) is given in terms of \textit{Whittaker}  $W-$ functions
\beq\label{mode_funct_A}
A_\pm(\tau,k)= \frac{1}{\sqrt{2k}} e^{\pm \pi \xi/2} W_{\pm i \xi , \frac{1}{2}} (-2 i k \tau)\,.
\eeq
The bare integrals in Eqs. \eqref{Energy formal} and \eqref{Helicity formal} can then be computed analytically by using the mode functions \eqref{mode_funct_A}, after imposing a comoving UV cut-off $\Lambda\, a(\tau)$ in order to identify the UV divergences.
We report in the following the final results for the bare energy density and helicity integrals evaluated in \cite{Ballardini:2019rqh}

\beq
\begin{split}
\frac{1}{2}\,\br\ve{E}^2 + \ve{B}^2\ke_\text{bare}\,=&\,\frac{\Lambda^4}{8\pi^2}+ \frac{ H^2 \Lambda^2 \xi^2}{8\pi^2}+ \frac{3 H^4\xi^2(5\xi^2-1)\log{(2 \Lambda/H)}}{16 \pi^2} \\
&+\frac{ H^4\xi^2  (-79 \xi^4 + 22\xi^2+29) }{64 \pi^2 (1+\xi^2)} + \frac{ H^4\xi (30 \xi^2-11) \sinh{(2\pi \xi)} }{64 \pi^3}\\
&+\frac{3 i H^4 \xi^2  (5 \xi^2-1) (\psi^{(1)}(1-i\xi)-\psi^{(1)}(1+i\xi))\sinh{(2\pi\xi)} }{64 \pi^3}\\
&- \frac{3 H^4 \xi^2  (5 \xi^2-1) (\psi(-1-i\xi)+\psi(-1+i\xi)) }{32 \pi^2}\,,
\end{split}\label{ED_bare}
\eeq

\beq
\begin{split}
\qquad\br\ve{E} \cdot \ve{B}\ke_\text{bare}=&\,
-\frac{ H^2  \Lambda ^2 \xi}{8 \pi ^2}
-\frac{3 H^4 \xi  \left(5\xi^2-1\right) \log \left(2 \Lambda/H\right)}{8 \pi ^2} 
\\
   &+\frac{ H^4 \xi (47\xi^2-22)}{16 \pi ^2}
   -\frac{H^4 (30\xi^2-11)\sinh{(2 \pi \xi)}}{32 \pi ^3} \\&-\frac{3 i H^4 \xi \left(5\xi^2-1\right)\le( \psi ^{(1)}(1-i \xi) -\psi ^{(1)}(1+i \xi )\ri) \sinh (2 \pi  \xi )}{32 \pi ^3}\\&+\frac{3 H^4  \xi  \left(5\xi^2-1\right)\le( \psi (1-i \xi) +\psi (1+i \xi )\ri)}{16 \pi ^2}\,,
\label{HE_bare}
\end{split}
\eeq
where $\p(x)$ is the Digamma function and $\p^{(1)}(x)\equiv {\rm d}\p(x)/{\rm d} x $.

These integrals, as expected for averaged quantities involving quadratic combinations of fields in curved space-times, show UV divergences. In particular we have quartic, quadratic and logarithmic UV divergences for the energy density of Eq. (\ref{ED_bare}), and only quadratic and logarithmic UV divergences for the helicity integral of Eq. (\ref{HE_bare}). On the other hand, they are well-behaved in the infrared, not exhibiting IR divergences.

\subsection{Adiabatic regularization}
To remove the UV divergences that affect the averaged energy density and helicity integrals of gauge fields, we subtract from the bare divergent quantities their respective adiabatic counterparts, following the procedure of adiabatic regularization highlighted above.

According to the standard convention, a mass regulator $m$ is added to the equation of motion  \eqref{eom_A}~\footnote{The mass term regulator $m$ is added to be in line with what is commonly done in literature performing adiabatic regularization for massless fields. However, it is worth noting that, according to our adiabatic procedure with IR cut-off, the addition of this mass regulator can be avoided \textit{ab initio}, being a redundancy in the regularization of the IR domain. Let us stress anyway that this easily leads us to extend our adiabatic results also to cases in which the physical mass of gauge fields is instead different from zero (see Appendix \ref{B}).}
\beq\label{eom_WKB}
\frac{{\rm d}^2}{{\rm d} \tau^2} A_\pm^\text{WKB}(\tau,k) + \left(k^2 \mp g k \phi'+\frac{m^2}{H^2 \tau^2}\right) A_\pm^\text{WKB}(\tau,k)=0\,,
\eeq
where the adiabatic mode function of gauge fields, for each polarization $\lambda=\pm$ is given by
\beq
A_\lambda^\text{WKB}(k,\tau)=\frac{1}{\sqrt{2 \Omega_\lambda(k,\tau)}} e^{-i \int \Omega_\lambda (k,\tau') {\rm d}\tau'}
\,.
\label{AwkbDefinition}
\eeq
Inserting this solution into the equation of motion \eqref{eom_WKB} we obtain the exact equation for the WKB frequency
\beq\label{WKB_eq}
\Omega_\lambda^2(k,\tau)= \bar{\Omega}^2_\lambda(k,\tau) + \frac{3}{4} \left(\frac{\Omega'_\lambda(k,\tau)}{\Omega_\lambda(k,\tau)}\right)^2-\frac{1}{2} \frac{\Omega''_\lambda(k,\tau)}{ \Omega_\lambda(k,\tau)}\,,
\eeq
where 
\beq
\bar{\Omega}_\lambda^2=\omega^2(k,\tau)-\lambda k g \phi'(\tau)\,, \qquad 
\omega^2(k,\tau)=k^2+m^2 a^2(\tau)\,.
\eeq

As described in the previous section, by solving \eqref{WKB_eq} iteratively, we can obtain the $n$-order adiabatic WKB frequencies. In the cases under consideration, the adiabatic expansion up to the fourth order is needed to remove the UV divergences. 

Thus, we obtain for the frequency up to the fourth adiabatic order 
\beq\label{adiab_freq}
\Omega_\lambda(k,\tau)= \bar{\Omega}_\lambda(k,\tau)+ \epsilon^2\, \Omega^{(2)}_\lambda(k,\tau)+ \epsilon^4\, \Omega^{(4)}_\lambda(k,\tau)\,,
\eeq
with, omitting for convenience the time and momentum dependence,
\beq
\Omega^{(2)}_\lambda=\frac{3}{8} \frac{(\bar{\Omega}^{\prime }_\lambda)^2}{\bar{\Omega}_\lambda^3}-\frac{1}{4}\frac{\bar{\Omega}_\lambda''}{\bar{\Omega}_\lambda^2}\,,
\eeq
and
\beq
\Omega^{(4)}_\lambda=-\frac{1}{2} \frac{(\Om_\lambda^{(2)})^2}{\bar{\Om}_\lambda}-\frac{3}{4}\frac{\Om_\lambda^{(2)} (\bar{\Om}_\lambda')^2}{\bar{\Om}_\lambda^4}+\frac{3}{4}\frac{\bar{\Om}_\lambda' \Om_\lambda^{(2)'}}{\bar{\Om}_\lambda^3}+\frac{1}{4}\frac{\Om_\lambda^{(2)} \bar{\Om}_\lambda''}{\bar{\Om}_\lambda^3}-\frac{1}{4}\frac{\Om_\lambda^{(2)''}}{\bar{\Om}_\lambda^2}\,,
\eeq
where we should further Taylor-expand $\bar{\Omega}_\lambda(k,\tau)$
in power of $\epsilon$ around ${\omega}(k,\tau)$ considering the term $(-\lambda k g \phi'(\tau))$ of adiabatic order one and discarding all the resulting terms of
adiabatic order larger than four in the final result.

We thus define the adiabatic mode functions in Eq. \eqref{AwkbDefinition} up to fourth order by using the adiabatic frequency in Eq. \eqref{adiab_freq}, and use them to compute the adiabatic counterparts of the energy density and helicity integrals by Eqs. \eqref{Energy formal} and \eqref{Helicity formal}.

We proceed now performing the adiabatic integrals, introducing the same comoving UV cut-off to regularize the UV-divergent terms.
Moreover, according to the proposed renormalization approach, a comoving IR cut-off $\b a(\t) H$ is considered, in order to take into account the fact that the adiabatic approximation breaks down at small wavenumbers.
In such a way we obtain the following results for the adiabatic counterparts of the energy density and helicity integrals (which match the ones in \cite{Ballardini:2019rqh} if also there $c=\beta a(\t) H$ is considered)
\beq
\begin{split}
\frac{1}{2}\,\br\ve{E}^2 + \ve{B}^2\ke _\text{ad}^{c=\beta H a(\t)}=&\,\,\frac{\Lambda ^4}{8 \pi ^2}+ \frac{H^2 \Lambda^2 \xi^2}{8\pi^2}+\frac{3 H^4 \xi^2 (5 \xi^2-1)  \log{(2 \Lambda/H)}}{16 \pi^2}\\
&-\frac{\beta ^4 H^4}{8 \pi ^2}- \frac{\beta^2 H^4 \xi^2}{8\pi^2}-\frac{3 H^4 \xi^2 (5 \xi^2-1) \log{(2 \beta)}}{16 \pi^2}\,,
\end{split}\label{ED_beta}
   \eeq
\beq
\begin{split}
\br\ve{E} \cdot \ve{B}\ke_\text{ad}^{c=\beta H a(\t)}=&\,-\frac{ H^2 \Lambda^2 \xi}{8 \pi^2}-\frac{3 H^4 \xi (5\xi^2-1) \log{(2\Lambda/H)}}{8 \pi^2} \quad \;\\
&+\frac{\beta^2  H^4 \xi}{8 \pi^2}+\frac{3 H^4 \xi (5\xi^2-1) \log{(2\beta)}}{8 \pi^2}\,.
\end{split}\label{HE_beta}
\eeq
We can immediately see that the terms proportional to the UV cut-off $\Lambda$ correctly reproduce the UV divergences of the bare quantities, so that, after subtraction, these infinities are removed.
 
The above expressions are obtained considering the $m\rightarrow0$ limit, which is indeed well-defined thanks to the introduction of the IR cut-off. Namely, no pathological dependencies on the mass term regulator $m$ manifest on the adiabatic results with a IR cut-off, as we can also see from the general case of $m\neq 0$ reported in Appendix \ref{B}.


\subsection{Matching the conformal anomaly}
In order to fix univocally the free parameter $\beta$, we can observe that in the conformal limit obtained for $m \rightarrow 0$ and $\xi \rightarrow 0$, 
the adiabatic expectation value of the energy density should provide the term connected to the conformal anomaly of gauge fields \cite{Chu:2016kwv, Dowker:1976zf, Brown:1977pq}. 
Indeed,  in the conformal limit, a proper renormalization scheme should provide the conformal anomaly induced by quantum effects \cite{Brown:1977pq, Capper:1974ic, Duff:1993wm, Duff:1977ay}
\beq 
\br T^\mu_{\phantom{m}\mu} \ke_\text{phys}=-\br T^\mu_{\phantom{\mu}\mu}\ke_\text{reg}\,,
\label{Eq.3.20}
\eeq
where $\br T^\mu_{\phantom{\mu}\mu} \ke_{\text{reg}}$ is the trace contribution to the energy-momentum tensor given by the particular renormalization method applied. 

In our case, since we are renormalizing physical quantities following the adiabatic subtraction, it follows that $\br T^\mu_{\phantom{\mu}\mu}\ke_\text{reg}=\br T^\mu_{\phantom{\mu}\mu}\ke_\text{ad}$.
Therefore, by requiring the right value of the conformal anomaly in the proper limit, the parameter $\beta$, which appears as a free parameter in the adiabatic expression of the energy density, can be fixed without ambiguities. Moreover, this physically motivated prescription for the value of the new degree of freedom immediately allows to obtain univocally defined finite results for the quantities of interest, after the subtraction has been performed.

In the particular case of conformally coupled massless gauge fields the expected value of the conformal anomaly should be twice the result of the conformal anomaly for a massless conformally coupled scalar field, namely $2 \times H^4/(960 \pi^2)= H^4/(480 \pi^2)$ \cite{Bunch:1978gb, Birrell:1982ix, Ballardini:2019rqh, Parker:2009uva}. This because the two helicities of the gauge fields are equivalent to two conformally coupled massless scalar fields for $\xi=0$.

By performing the $m \rightarrow 0$ and $\xi \rightarrow 0$ limits for the case of gauge fields we obtain
\beq
\lim_{\xi\rightarrow 0,\, m \rightarrow 0} \br T^\mu_{\phantom{\mu}\mu}\ke_\text{ad}=\lim_{\xi\rightarrow0,\ m \rightarrow 0}\frac{ \br\ve{E}^2 + \ve{B}^2\ke _{\text{ad}}^{c=\beta H a(\t)}}{2}=-\frac{\beta^4 H^4}{8 \pi^2}\,,
\eeq
and this term should reproduce the expected value of the trace anomaly. Accordingly to \eqref{Eq.3.20}, the matching procedure gives
\beq
\frac{\beta^4 H^4}{8 \pi^2}= \frac{H^4}{480 \pi^2} \implies \beta = \frac{1}{\sqrt{2}\times 15^{1/4}} \approx 0.359\,.
\label{fixed beta}
\eeq

Therefore, we have a physically motivated prescription that is able to fix univocally the renormalization scheme. As a consequence, after the adiabatic subtraction is performed, we are able to obtain univocal finite results for the averaged energy density and helicity of gauge fields. Most importantly, our adiabatic renormalization method succeeds in providing the conformal anomaly in the proper limit, where instead the standard adiabatic procedure fails, leading to pathological infrared divergences.

Let us finally add a further remark. According to our proposal, it is clear that the IR cut-off should be introduced in any case when performing adiabatic renormalization, for the reasons  explained in Sec. \ref{sezione3} . Here we show how the use of this comoving cut-off does not spoil required results already obtained within the standard approach, as one would expect.
To this purpose, let us consider the standard case of a conformally coupled massless scalar field, where no IR divergences appear in the usual adiabatic renormalization procedure of the energy-momentum tensor. The standard adiabatic subtraction seems to be not problematic, indeed one is able to obtain the conformal anomaly, one of the main results required for a renormalization procedure.
However, by explicit calculation we obtain no dependence on the new IR cut-off within our approach, so we are able to obtain the well-known result for the conformal anomaly 
\beq 
\begin{split}
\br T^\mu_{\phantom{\mu}\mu} \ke_\text{phys}=&-\br T^\mu_{\phantom{\mu}\mu} \ke_\text{ad}^{c=\beta H a(t)}=\\
&-\frac{\ap(t)^2 \app(t)}{160 \pi ^2 a(t)^3}+\frac{\app(t)^2}{480 \pi ^2 a(t)^2}+\frac{\ap(t) \dddot{a}(t)}{160 \pi ^2 a(t)^2}+\frac{\ddddot{a}(t)}{480 \pi ^2 a(t)}\,.
\end{split}
\eeq
This simple example shows how the adiabatic procedure here introduced produces the correct result also in this case. The investigation of the impact of this renormalization scheme on observables evaluated in other inflationary models will be the subject of future works.

\subsection{Renormalized results and comparison with minimal subtraction scheme}
We can now perform the proper renormalization procedure, by subtracting to the bare results of the energy density in Eq. \eqref{ED_bare} and of the helicity integral in Eq. \eqref{HE_bare} their adiabatic counterparts in Eqs. \eqref{ED_beta} and \eqref{HE_beta} respectively, where $\beta$ has to be fixed according to Eq. \eqref{fixed beta}.
The final renormalized result for the energy density is thus
\beq
\label{energybetascheme}
\begin{split}
\frac{1}{2}\,\br\ve{E}^2 + \ve{B}^2\ke_{\b}=&\, \frac{2 H^4}{960 \pi^2} +\frac{H^4 \xi^2\left(-1185 \xi^4 + (330+4\sqrt{15})\xi^2+435+4\sqrt{15}\right) }{960 \pi ^2 \left(1+\xi ^2\right)}\\
   &-\frac{3 H^4 \xi ^2   \left(5 \xi ^2-1\right) \log \left(15/4\right)}{64 \pi ^2}+\frac{ H^4\xi    \left(30 \xi
   ^2-11\right) \sinh (2 \pi  \xi )}{64 \pi ^3}\\
   &-\frac{3  H^4  \xi^2\left(5 \xi ^2-1\right) (\psi ^{(0)}(-1-i \xi )+\psi ^{(0)}(-1+i \xi ))}{32 \pi ^2}\\
   &+\frac{3 i  H^4 \xi^2
   \left(5 \xi ^2-1\right) (\psi ^{(1)}(1-i \xi )-\psi ^{(1)}(1+i \xi )) \sinh (2 \pi  \xi )}{64 \pi ^3}\,,
   \end{split}
   \eeq
and the one for the helicity integral is 
\beq\label{helicitybetascheme}
\begin{split}
\br\ve{E} \cdot \ve{B}\ke_{\b}= &\,\frac{H^4 \xi  \left(705 \xi^2-330 -\sqrt{15}\right)}{240 \pi ^2}+\frac{3 H^4 \xi  \left(5 \xi ^2-1\right) \log \left(15/4\right)}{32 \pi ^2}\\
&+\frac{3 H^4 \xi  \left(5 \xi ^2-1\right) (\psi ^{(0)}(1-i \xi )+\psi ^{(0)}(1+i \xi ))}{16 \pi ^2}\\
&+\frac{3 i H^4 \xi  \left(5 \xi ^2-1\right) (-\psi ^{(1)}(1-i \xi )+\psi ^{(1)}(1+i \xi )) \sinh (2 \pi  \xi )}{32 \pi ^3}\\
&+\frac{H^4 \left(11-30 \xi ^2\right) \sinh (2 \pi  \xi )}{32 \pi ^3}\,.
\end{split}
\eeq

It is instructive to compare the above results with the ones obtained by a minimal subtraction (MS) scheme, where only the UV divergences are removed (as in \cite{Ballardini:2019rqh})

\beq
\label{energyMSscheme}
\begin{split}
\frac{1}{2}\,\br\ve{E}^2 + \ve{B}^2\ke_\text{MS}=&\,\frac{ H^4\xi^2  (-79 \xi^4 + 22\xi^2+29) }{64 \pi^2 (1+\xi^2)} + \frac{ H^4\xi (30 \xi^2-11) \sinh{(2\pi \xi)} }{64 \pi^3}\\
&+\frac{3 i H^4 \xi^2  (5 \xi^2-1) (\psi^{(1)}(1-i\xi)-\psi^{(1)}(1+i\xi))\sinh{(2\pi\xi)} }{64 \pi^3}\\
&- \frac{3 H^4 \xi^2  (5 \xi^2-1) (\psi(-1-i\xi)+\psi(-1+i\xi)) }{32 \pi^2}\,,
   \end{split}
\eeq

\beq\label{helicityMSscheme}
\begin{split}
\br\ve{E} \cdot \ve{B}\ke_\text{MS}=&+\frac{ H^4 \xi (47\xi^2-22)}{16 \pi ^2}
   -\frac{H^4 (30\xi^2-11)\sinh{(2 \pi \xi)}}{32 \pi ^3} \\&-\frac{3 i H^4 \xi \left(5\xi^2-1\right)\le( \psi ^{(1)}(1-i \xi) -\psi ^{(1)}(1+i \xi )\ri) \sinh (2 \pi  \xi )}{32 \pi ^3}\\&+\frac{3 H^4  \xi  \left(5\xi^2-1\right)\le( \psi (1-i \xi) +\psi (1+i \xi )\ri)}{16 \pi ^2}\,.
\end{split}
\eeq

\begin{figure}[H]
\centering
\includegraphics[width=\textwidth]{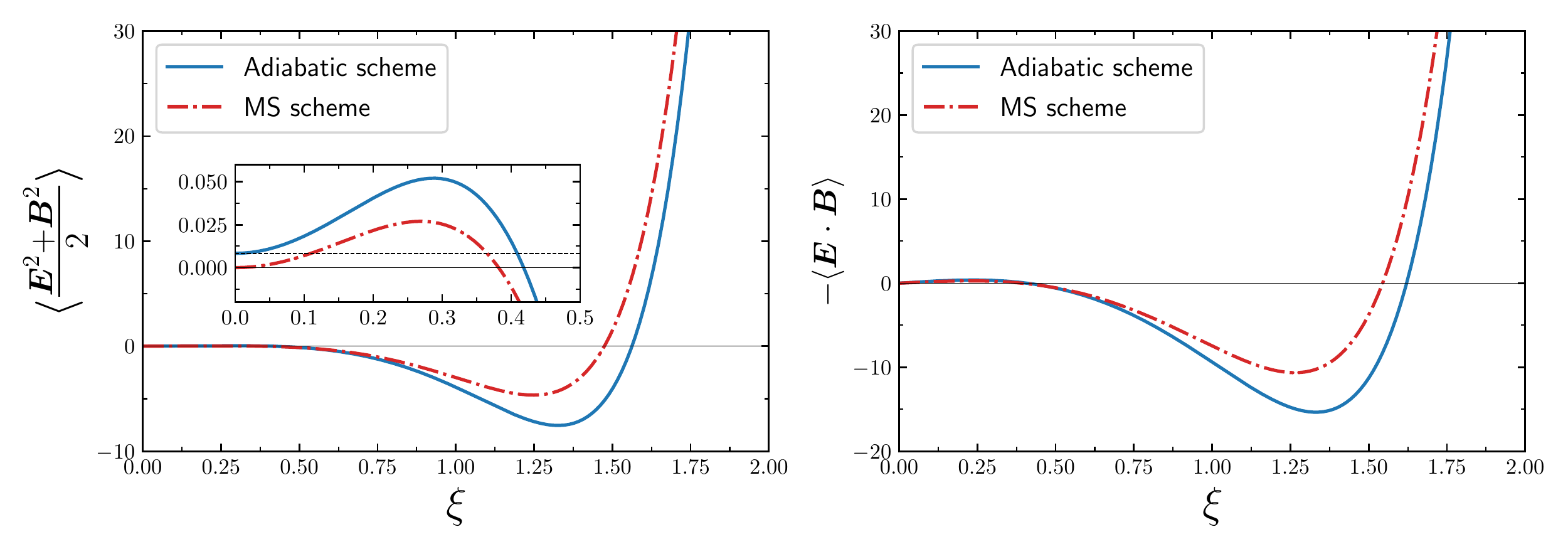}
\caption{We compare the adiabatic and MS renormalization schemes for the energy density (left panel) and for the helicity integral (right panel). The inset on the left panel shows the behavior of the energy density for $\xi$ close to zero. Notice that only the adiabatic scheme reproduces the correct value of the conformal anomaly (horizontal dashed line).}\label{Fig.1}
\end{figure}
In Fig. \ref{Fig.1} we plot the renormalized energy density in the adiabatic \eqref{energybetascheme} and MS \eqref{energyMSscheme} scheme (left panel) and, similarly, the renormalized helicity integral in the adiabatic \eqref{helicitybetascheme} and MS \eqref{helicityMSscheme} scheme (right panel) in units of $(2 \pi)^2/H^4$. 
As one can see, the differences between the two schemes are of order $\mathcal{O}(1)$ at small couplings. The leading asymptotic behaviors for $\xi \gg 1$ in the adiabatic scheme are given by 
\beq
\frac{1}{2}\,\br\ve{E}^2 + \ve{B}^2\ke_\beta \sim \frac{9 H^4 \sinh(2 \pi \xi)}{1120 \pi^3 \xi^3}\,,\qquad -\br\ve{E} \cdot \ve{B}\ke_\beta \sim \frac{9 H^4 \sinh(2 \pi \xi)}{560 \pi^3 \xi^4}\,,
\eeq
and reproduce, as expected, the asymptotic behaviors in the MS scheme \cite{Ballardini:2019rqh}. The adiabatic subtraction introduces further power law corrections that are however subleading.
As expected, the inset on the left panel shows that only the adiabatic scheme provides the correct value of the conformal anomaly for $\xi \to 0$. 

We would like to conclude this section bringing the attention to the fact that the adiabatic procedure is not just a tool to identify the UV-divergent terms of quantities involving expectation values of quantum fields in curved space-times, but it is a proper renormalization prescription. This means, in particular, that it should provide univocally defined finite results. Moreover, it is properly the subtraction of the contribution of the adiabatic vacuum that introduces surprising effects in the renormalization of physical quantities. A remarkable example is the generation of conformal anomalies which break the classical conformal symmetry at the quantum level and that accompany the renormalization of the energy-momentum tensor of conformally coupled fields in cosmological space-times. This last aspect notably underlines how the totality of the adiabatic counterpart has to be subtracted to the bare divergent quantities. As pointed out in the introduction, there are cases in which the standard adiabatic regularization leads to not well-defined adiabatic terms in the infrared regime, forcing to fix a different prescription, as the minimal scheme, for the subtraction, loosing the uniqueness of the finite term. In particular, as shown in the analyzed case of gauge fields, the conformal anomaly was not recovered within the minimal scheme.
This further reinforces the idea that such minimal subtraction is not the proper prescription for adiabatic regularization.


\newpage

\section{Discussion and conclusions}
\label{sezione Conclusioni}

In this manuscript we rediscussed the adiabatic renormalization in curved space-time. In line with previous literature, we questioned about the correct use of the adiabatic subtraction and, by examining its foundations and its essential properties, we showed how its range of validity has to be restricted to the UV regime. Consequently we evinced how the adiabatic subtraction should be considered only for modes up to a scale comparable with the Hubble horizon. This not only fully meets the essence of the adiabatic approximation, but is also needed, otherwise problematic features can manifest in the deep IR, which could be plagued by unphysical divergences.
Furthermore, we pointed out that, otherwise, in order to extend the adiabatic renormalization up to the IR, one should take into account all the adiabatic orders for the subtraction and not only those that ensures the removal of the UV infinities, at the price of loosing the 
predictivity of the adiabatic approach.

Accordingly, we suggested to supplement the adiabatic renormalization frame with the introduction of a comoving IR cut-off of the form $c=\beta a H$, which stops the adiabatic subtraction approximatively around the horizon crossing of modes. The new parameter $\beta$ has then to be fixed by a suitable physical prescription, making the renormalization scheme univocally defined.
In particular we emphasized that, as part of a well-defined renormalization scheme, the conformal anomaly for massless conformally coupled (quantum) fields in curved space-times should be always reproduced.
Let us further underline how this request lets no ambiguities in the choice of the infrared cut-off for the adiabatic counterpart and returns us unique renormalized physical quantities.

In light of this perspective, we applied this new approach to the particular and considerable case of a $U(1)$ gauge field coupled to an axion-like inflaton, which is specially suitable to show how our procedure is able to provide a well-defined renormalized energy-momentum tensor also in the IR, where the standard adiabatic procedure failed. In this case, using the fact that the conformal anomaly is $2 \times H^4/(960 \pi^2)= H^4/(480 \pi^2)$, since the two physical states $A_{\pm}$ of the gauge field are equivalent to two conformally coupled massless scalar fields for $\xi=0$, we obtained the value for the new degree of freedom introduced by the IR cut-off $\beta=1/(\sqrt{2}\times 15^{1/4}) \approx 0.359$, fixing univocally the finite part and thus the renormalization scheme. 
 
Moreover, this new procedure of adiabatic renormalization is
also needed for the study of the backreaction of gauge fields on the axion-like (inflaton) evolution, due also to the contribution of the helicity integral in the equation of motion of the inflaton field. This backreaction impacts on the dynamics of the inflaton, possibly changing the duration of the inflationary phase. We hope to investigate such aspects in a future work.

 Let us finally recall that the need of a physical prescription to fix univocally the renormalization scheme is a fundamental ingredient of all renormalization schemes also in flat space-time quantum field theories. In particular, since IR divergences are not universal as the UV ones, but depend on the choice of the state of reference, this seems to indicate an interesting link between our adiabatic renormalization prescription and algebraic renormalization techniques, where the choice among the possible vacuum Hadamard states is fixed at the level of physical observables (see \cite{Hollands:2014eia, Fredenhagen:2014lda, Brunetti:2009pn} for a  review on the subject).

Within this work we tried to underline the necessity of a new prescription for the adiabatic renormalization scheme. As shown, this has strong effects in particular for such models where otherwise the standard adiabatic approach fails, producing unphysical IR divergences. The latter is the case for the model presented here of an axion-like inflaton coupled to U(1)-gauge fields, but is also the case for all other models with the same properties in the IR (see e.g. \cite{Lozanov:2018kpk, Ishiwata:2021yne, Kamada:2020jaf, Hashiba:2021gmn} ). We plan to apply our renormalization procedure for those particular models in future works.

\acknowledgments
The authors are thankful to Matteo Braglia and Nicola Pinamonti for useful discussions and correspondence. The authors are supported in part by INFN under the program TAsP ({\it Theoretical Astroparticle Physics}).

\appendix
 
\section{Conservation of $T_{\mu\nu}$} \label{A}
According to \cite{Bastero-Gil:2013nja}, performing the adiabatic subtraction only over high momentum modes by the introduction of  a time-dependent cut-off, leads to the violation of the energy-momentum covariant conservation by changing the time derivative of the energy density (or, more in general, of the $T_{\mu\nu}$ components).
Here we show how this is not true within our approach, which therefore leads to a renormalized energy-momentum tensor that does not spoil any conservation.

As an illustrative example, let us consider a generic expression for the adiabatic counterpart of the energy density $\r$ calculated using a comoving IR cut-off. In a complete general way this quantity can be expressed as
\beq 
\r = \int_{\b a(t) H(t)} {\rm d}k\, \frac{k^n}{(a(t) H(t))^{n+1}} \, g\left(\frac{k} {a(t) H(t)} , t\right) \,,
\eeq
with $g$ an adimensional function of $k/(a(t) H(t))$ and of the proper time $t$. 
It is easy to see that the introduction of a comoving IR cut-off proportional to $a(t) H(t)$ does not change the time dependence of the integral. Indeed, setting $x=k/(a(t) H(t))$ the integral becomes
\beq
\r = \int_{\b} {\rm d}x\, x^n g(x, t) \,,
\eeq
and performing a time derivative one obtains
\begin{equation}
\dot{\r} = \int_{\b} {\rm d}x\, x^n \frac{{\rm d}}{{\rm d} t}g(x, t) \,.
\end{equation} 
 We can therefore conclude that the criticism ilustrated in \cite{Bastero-Gil:2013nja} does not apply in our case, since the apparent time dependence of the introduced cut-off is an artefact of the comoving coordinate choice, which can be removed by rewriting all in terms of the physical wavenumber scaled by the Hubble factor $k/(a(t) H(t))$. To conclude, if for $T_{\mu\nu}$ there is conservation mode-by-mode, namely for each $k$ and $t$, than there will be also for each $x$ and $t$ and the presence of the comoving IR cut-off will not affect such conservation.

 \section{Adiabatic results for U(1)-gauge massive fields}\label{B}
In Sec. \ref{sezione3} we focused on the case of massless gauge fields, both because it has a pathological behavior in the IR within the standard adiabatic procedure, making particularly evident the usefulness of our method, and because it is of phenomenological interest. 
In any case, our procedure should be applied in general, hence also to the massive gauge field case (see e.g. \cite{Maleknejad:2012fw} for some applications). 

We report here the full results for the adiabatic integrals of the energy density and helicity extended to the case of massive gauge fields.\footnote{Notice that in the massive case the integral in Eq. \eqref{Energy formal} becomes \beq
\frac{\br\ve{E}^2+\ve{B}^2\ke}{2}=\int \frac{{\rm d}k}{(2\pi)^2 a(\tau)^4} k^2 \left[|A'_+|^2+|A'_-|^2+\le(k^2 + m^2 a(\t)^2 \ri)\le(|A_+|^2+|A_-|^2\ri)\right]\,.
\eeq.}

\beq
\begin{split}
\frac{1}{2}\,\br\ve{E^2 + B^2}\ke _{\text{ad},m}^{c=\beta H a(\t)}=\,&\frac{\Lambda ^4}{8 \pi ^2}+\frac{\Lambda ^2 \left(m^2+ H^2 \xi ^2\right)}{8 \pi ^2}\\
&-\frac{\left(m^4+6 H^2 m^2 \xi ^2+3 H^4 \xi ^2 \left(1-5 \xi ^2\right)\right) \log
   \left(2 \Lambda/H\right)}{16 \pi ^2}\\
   &+\frac{15 m^4+20 H^2 m^2 \left(1+15 \xi ^2\right)-2 H^4 \left(1-270 \xi ^2+690 \xi ^4\right)}{960 \pi ^2}\\
   &-\frac{H^{13} \beta ^9 \left(-1+60 \beta ^4+270 \xi ^2+60 \beta ^2 \xi ^2-690 \xi ^4\right)}{480 \pi ^2 \left(m^2+H^2 \beta ^2\right)^{9/2}}\\
   &-\frac{H^{11} m^2 \beta ^7 \left(660 \beta ^4+20 \beta ^2 \left(1+42 \xi ^2\right)-9 \left(1-150 \xi ^2+540 \xi ^4\right)\right)}{960 \pi ^2 \left(m^2+H^2 \beta
   ^2\right)^{9/2}}\\
   &-\frac{H^9 m^4 \beta ^5 \left(27+1000 \beta ^4+940 \xi ^2-4320 \xi ^4+40 \beta ^2 \left(1+36 \xi ^2\right)\right)}{640 \pi ^2 \left(m^2+H^2 \beta ^2\right)^{9/2}}\\
   &-\frac{H^7 m^6 \beta ^3 \left(1+480 \beta ^4+208 \xi ^2-1040 \xi ^4+16 \beta ^2 \left(1+44 \xi ^2\right)\right)}{256 \pi ^2 \left(m^2+H^2 \beta ^2\right)^{9/2}}\\
   &-\frac{H^5 m^8 \beta  \left(60 \beta ^4+9 \xi ^2-45 \xi ^4+\beta ^2 \left(1+78 \xi ^2\right)\right)}{48 \pi ^2 \left(m^2+H^2 \beta ^2\right)^{9/2}}\\
   &-\frac{H^3 m^{10} \left(7 \beta ^3+6 \beta  \xi ^2\right)}{16 \pi ^2 \left(m^2+H^2 \beta ^2\right)^{9/2}}-\frac{H m^{12} \beta }{16 \pi ^2 \left(m^2+H^2 \beta ^2\right)^{9/2}}\\
   &+\frac{\left(m^4+6 H^2 m^2 \xi ^2+3 H^4 \xi ^2 \left(1-5 \xi ^2\right)\right) \log \left(\beta + (\sqrt{m^2+H^2\beta^2})/H\right)}{16 \pi ^2}\,,
   \end{split}
\eeq
\beq
\begin{split}
\qquad \qquad \br\ve{E \cdot B}\ke _{\text{ad},m}^{c=\beta H a(\t)}=&-\frac{H^2 \Lambda ^2 \xi }{8 \pi ^2}+\frac{3 H^2 \xi  \left(3 m^2+H^2 \left(1-5 \xi ^2\right)\right) \log \left(2\Lambda/H\right)}{8 \pi ^2}\\
&+\frac{3 H^2 \xi  \left(-3 m^2+H^2 \left(-1+5 \xi ^2\right)\right) \log \left(\beta + (\sqrt{m^2+H^2\beta^2})/H\right)}{8 \pi ^2}\\
&+\frac{H^2 \xi  \left(-21 m^2+H^2 \left(-19+56 \xi ^2\right)\right)}{16 \pi ^2}
+\frac{H^{13} \beta ^9 \xi  \left(19+2 \beta ^2-56 \xi ^2\right)}{16 \pi ^2 \left(m^2+H^2 \beta ^2\right)^{9/2}}\\
&+\frac{H^{11} m^2 \beta ^7 \xi  \left(49+60 \beta ^2-344 \xi ^2\right)}{32 \pi ^2 \left(m^2+H^2 \beta ^2\right)^{9/2}}
+\frac{H^9 m^4 \beta ^5 \xi  \left(43+96 \beta ^2-216 \xi ^2\right)}{16 \pi ^2 \left(m^2+H^2 \beta ^2\right)^{9/2}}\\
&+\frac{H^7 m^6 \beta ^3 \xi  \left(13+64 \beta ^2-65 \xi ^2\right)}{8 \pi ^2 \left(m^2+H^2 \beta ^2\right)^{9/2}}
+\frac{9 H^3 m^{10} \beta  \xi }{8 \pi ^2 \left(m^2+H^2 \beta ^2\right)^{9/2}}\\
&+\frac{3 H^5 m^8 \beta  \xi  \left(1+13 \beta ^2-5 \xi ^2\right)}{8 \pi ^2 \left(m^2+H^2 \beta ^2\right)^{9/2}}\,.
\end{split}
\eeq
Naturally,  the parameter $\b$ is still fixed by  Eq. \eqref{fixed beta} , due to the physical property that for $\xi$ and $m \ra 0$ one should match the conformal anomaly.

\newpage

\bibliographystyle{JHEP}
\bibliography{BIBLIO}

\end{document}